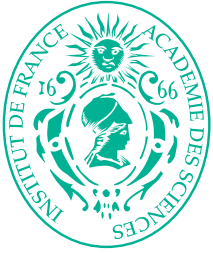

INSTITUT DE FRANCE
Académie des sciences

# *Comptes Rendus*

## *Physique*

Gael Chauvin

**Direct imaging of exoplanets: Legacy and prospects**

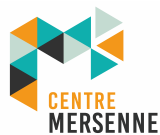

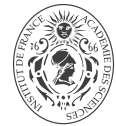

Exoplanets / *Exoplanètes*

# Direct imaging of exoplanets: Legacy and prospects

*Imagerie directe des exoplanètes : Héritage et perspectives*

Gael Chauvin [a]

[a] J.-L. Lagrange Laboratory , Côte d'Azur Observatory, UMR 7293 Bâtiment H. Fizeau, 28, Avenue Valrose, Nice cedex 2, 06108, France

*E-mail:* gael.chauvin@oca.eu

**Abstract.** Understanding how giant and terrestrial planets form and evolve, what is their internal structure and that of their atmosphere, represents one of the major challenges of modern astronomy, which is directly connected to the ultimate search for life at the horizon 2030–2050. However, several astrophysical (understanding of the formation and physics of giant and terrestrial exoplanets), biological (identification of the best biomarkers) and technological (technical innovations for the new generations of telescopes and instruments) obstacles must be overcome. From the astrophysical point of view, it is indeed crucial to understand the mechanisms of formation and evolution of giant planets, including planet and disk interactions, which will completely sculpt the planetary architectures and thus dominate the formation of terrestrial planets, especially in regions around the host star capable of supporting life. It is also important to develop dedicated instrumentation and techniques to study in their totality the population of giant and terrestrial planets, but also to reveal in the near future the first biological markers of life in the atmospheres of terrestrial planets. In that perspective, direct imaging from ground-based observatories or in space is playing a central role in concert with other observing techniques. In this paper, I will briefly review the genesis of this observing technique, the main instrumental innovation and challenges, stellar targets and surveys, to then present the main results obtained so far about the physics and the mechanisms of formation and evolution of young giant planets and planetary system architectures. I will then present the exciting perspectives offered by the upcoming generation of planet imagers about to come online, particularly on the future extremely large telescopes. On the timescale of a human Life, we may well be witnessing the first discovery of an exoplanet and the first detection of indices of life in the atmosphere of a nearby exo-Earth!

**Résumé.** Comprendre comment les planètes géantes et terrestres se forment et évoluent, quelle est leur structure interne et celle de leur atmosphère, représente l'un des défis majeurs de l'astronomie moderne, qui est directement lié à la recherche ultime de la vie à l'horizon 2030–2050. Cependant, plusieurs obstacles astrophysiques (compréhension de la formation et de la physique des exoplanètes géantes et terrestres), biologiques (identification des meilleurs biomarqueurs) et technologiques (innovations techniques pour les nouvelles générations de télescopes et d'instruments) doivent être surmontés. Du point de vue astrophysique, il est en effet crucial de comprendre les mécanismes de formation et d'évolution des planètes géantes, y compris les interactions entre la planète et le disque, qui vont complètement sculpter les architectures planétaires et ainsi dominer la formation de planètes terrestres, notamment dans les régions autour de l'étoile hôte capables d'accueillir la vie. Il est également important de développer une instrumentation et des techniques dédiées pour étudier dans leur totalité la population de planètes géantes et terrestres, mais aussi de révéler dans un futur proche les premiers marqueurs biologiques de la vie dans les atmosphères des planètes terrestres. Dans cette perspective, l'imagerie directe depuis des observatoires au sol ou dans l'espace joue un rôle central, de concert avec d'autres techniques d'observation. Dans cet article, je rappellerai brièvement la genèse de cette technique d'observation, les principales innovations et défis instrumentaux, les cibles stellaires et les relevés, pour ensuite présenter les principaux résultats obtenus jusqu'à présent sur la physique et les mécanismes de formation et dévolution des jeunes planètes géantes et les architectures des systèmes planétaires. Je présenterai ensuite les perspectives passionnantes offertes par la prochaine génération d'imageurs de planètes sur le point d'être mis en ligne, notamment sur les futurs extrêmement grands télescopes. A l'échelle d'une vie humaine, nous pourrions bien assister à la première découverte d'une exoplanète et à la première détection d'indices de vie dans l'atmosphère d'une exo-Terre proche !

## 1. Introduction

### 1.1. *A rich legacy*

Today's success of direct imaging of exoplanets and planetary systems is intimately connected to the pioneer work in the late 80's and early 90's dedicated to the development of Adaptive Optics (AO) system, infrared (IR) detectors, and coronographic techniques for the instrumentation of ground-based telescopes. Precursor instruments on 4-m class telescopes rapidly demonstrated performances that could compete with space as illustrated by the emblematic discoveries and images of the ultracool brown dwarf companion Gl 229 B [1] and the edge-on circumstellar disk around $\beta$ Pictoris [2]. They soon motivated the research and development of more sophisticated AO planet imagers on 10 m-class telescopes of first and second generations, now routinely operating on world-class observatories, and preparing the path forward the future planet imagers of the extremely large telescopes in the decade to come. Joining the small family of planet hunting techniques with radial velocity, transit, $\mu$-lensing and astrometry, direct imaging (and interferometry, see [3]) offers a unique opportunity to explore the outer part of exoplanetary systems beyond 5–10 au to complete our view of planetary architectures, and to explore the properties of relatively cool self-luminous giant planets (see Figure 1). The exoplanet's light can indeed be spatially resolved and dispersed to probe the atmospheric properties of exoplanets (and brown dwarf companions). As today's imaged exoplanets are young (because they are hotter, brighter, thus easier to detect than their older counterparts), their atmospheres show low-gravity features, as well as the presence of clouds, and non-equilibrium chemistry processes. These physical conditions are very different and complementary to the ones observed in the atmospheres of field brown dwarfs or irradiated inflated Hot Jupiters (studied in spectroscopy or with transit techniques like transmission and secondary-eclipse, respectively). Finally, direct imaging enables to directly probe the presence of planets in their birth environment. Planet characteristics and disk spatial structures can then be linked to study the planet–disk interactions and the planetary system's formation, evolution, and stability, which is a fundamental and necessary path to understand the formation of smaller terrestrial planets with suitable conditions to host life. After briefly describing the key components of the current planet imagers, and the main characteristics of the nearby stars observed, I will summarize the key results obtained so far and main perspectives offered for the decade to come.

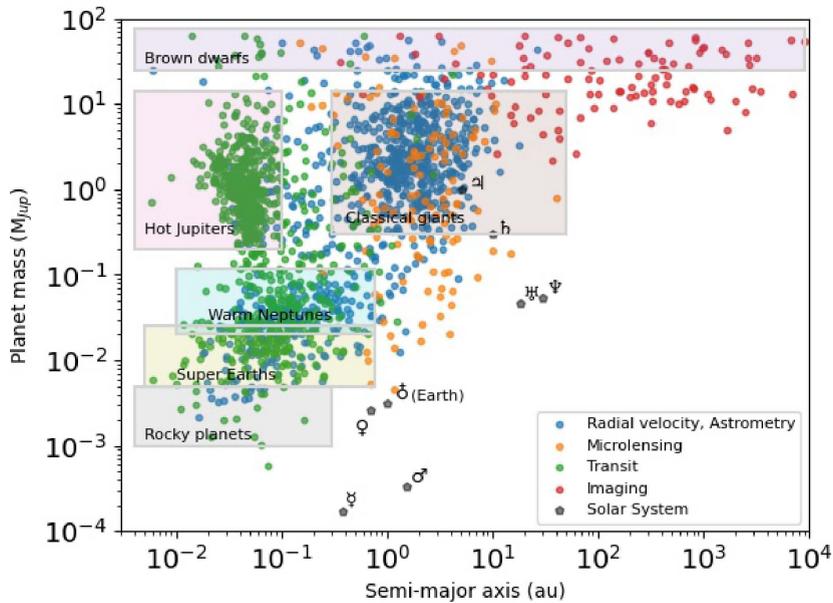

**Figure 1.** Exoplanets (and brown dwarf companions) discoveries considering the different planet hunting techniques (from http://exoplanet.eu). The exoplanets masses are reported as a function of their distances to stars (with minimum masses for the radial velocity method). The different categories of exoplanets are reported including the rocky planets, super-Earths, warm Neptunes, Hot Jupiter and classical giants. The brown dwarf companions are also reported as both populations probably overlap although a typical limit at 20–30 Jupiter masses is generally adopted given the observed companion mass distribution to stars.

## 1.2. *Instrumental innovation and challenges*

The success of direct imaging relies on a sophisticated instrumentation designed to detect faint planetary signals angularly close to a bright star. A Jupiter-like planet (orbiting at 5 au) around a typical young, nearby star at 50 pc would lie at an angular separation of 100 mas setting the order of angular separation we aim at. The typical planet–star contrast are about $10^{-6}$ for a young Jupiter (today's performances) and $10^{-8}$ for a mature Jupiter observed in emitted light. It goes down $10^{-8}$–$10^{-9}$ for a super-Earth in reflected light. From the early discoveries of young, massive self-luminous giant planets like $\beta$ Pic b (*H*-band contrast of $10^{-4}$ at 200–400 mas) with the first generation of planet imagers at the Very Large Telescope (VLT), and the Palomar, Subaru, Keck, and Gemini observatories to the first images of an Exo-Earth with maybe an Extremely Large Telescope (ELT), several orders of magnitudes in contrast must be covered. Therefore, innovative technological developments are required to meet the ultimate goal of imaging Exo-Earths (including the construction of extremely large telescope and the ability of achieving high-quality wavefront control combined with ultimate coronographic and differential techniques). In that perspective, the second generation of planet imagers, like the European Spectro-Polarimetric High-Contrast Exoplanet Resarch instrument [4] (SPHERE, see Figure 2 (Left), the Gemini Planet Imager [5] (GPI), and the Subaru Coronagraphic Extreme Adaptive Optics (SCExAO) instrument, aim at pushing high-contrast performances down to $10^{-6}$–$10^{-7}$ at typical separation of 200 mas. They validate a dedicated instrumentation based on a 3-stages implementation design with: (i) high angular resolution access, (ii) stellar light attenuation using coronography, (iii) speckle subtraction using differential imaging techniques. In the last step can be added the recent development of powerful post-processing tools to optimize the stellar signal suppression. These instruments already serve as references and demonstrators for future terrestrial planet imagers of the ELTs, like PCS [6] for the European ELT.

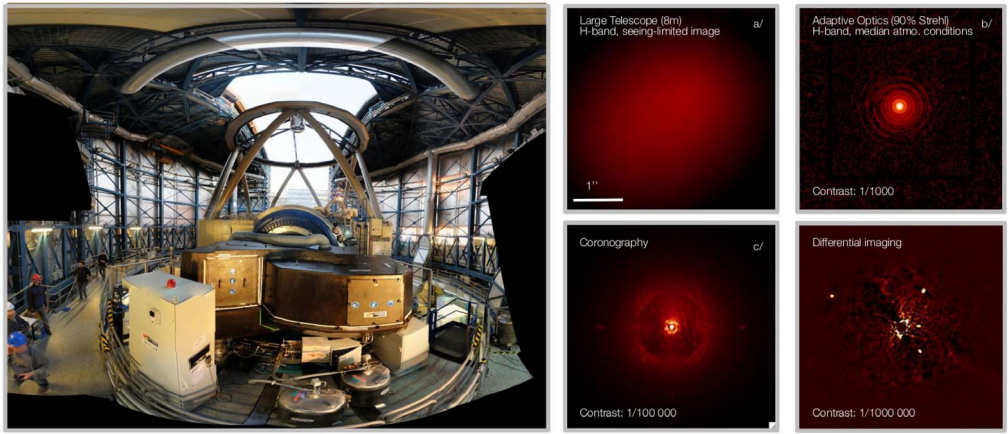

**Figure 2.** Left, the SPHERE planet imager at UT3/VLT. Right, high angular resolution and high contrast implementation three main steps for direct imaging of exoplanets including XAO (a/), coronography (b/) and differential imaging including post-processing (c/) to optimize the subtraction of the stellar residual while conserving the exoplanet signal.

From the ground, the atmosphere turbulence affects the light propagation and prevents large telescopes from reaching a spatial resolution at the diffraction limit. Current Extreme-Adaptive Optics (XAO) systems enable to compensate for atmospheric, but also telescope and common path defects to routinely reach nowadays Strehl[1] correction of 90% in H-band on bright ($V \leq$ 10 mag) targets. XAO-instruments like SPHERE, GPI, and SCExAO rely on high-order deformable mirror, fast-temporal sampling frequency at more than 1.0 kHz, spatial filtering of the wavefront before sensing, tuned calibrations of the instrumental defects and enhanced stability to limit low-order wavefront errors and alignment drift. The main specifications are driven by the need to get (i) optimal spatial sampling of the wavefront given the turbulence coherence length (sampling better than $(D/r_0)^2$; $r_0 \sim$ 20 cm at Paranal) and (ii) optimal temporal sampling to beat the turbulence speed (coherence time $\tau_0$; 2–5 ms at Paranal) to minimize the errors of wavefront reconstruction and command. The correction stability in terms of Strehl-correction, control of low-order aberrations and pointing stability are essential to avoid any leakage or PSF deformation during the observation that would significantly degrade the planet detection performances.

With a stable and diffraction-limited Point-Spread Function (PSF), the motivation for coronography is relatively intuitive as it consists in blocking the light from the central star to search for fainter objects in the close stellar environment. Simple occulting masks or classical Lyot coronographs have been massively used in the past years on high-contrast imagers to essentially: (i) reduce by a typical factor of ~100 the intensity of the central star diffracted-limited core and avoid any saturation effects, (ii) reduce the intensity of the PSF wings without canceling the off-axis planetary signal, (iii) improve the observing efficiency by reducing readout overheads, and (iv) reduce the total read-out noise. Classical and more recent apodized versions of the Lyot coronagraph are typically limited to 3–4$\lambda/D$ inner working angle (IWA),[2] mostly because they rely on amplitude manipulation to attenuate starlight diffraction, covering a large area of the PSF at the coronagraph plane. With the improvement of instrumental stability achieved with the new generation fo planet imagers, new concepts have emerged, particularly to access smaller inner working angles down to $\lambda/D$ (see apodized vortex and phase-induced amplitude coronographs). Pushing the instrumental limitation to access smaller IWA down to $\lambda/D$ or 2$\lambda/D$ (40 to 80 mas for the VLT at $H$-band) can make a significant scientific breakthrough as the bulk of the giant planet population around young, nearby stars is predicted to be located beyond the ice-line at ~3 au (i.e. 100 mas for a star at 30 pc).

---
[1] The Strehl ratio can be defined as the ratio between the peak of the AO corrected PSF and the one of the ideal diffraction-limited Airy pattern.

As shown in Figure 2 (Right), boiling atmospheric speckles and instrumental quasi-statics speckles still remain an important source of limitation in the XAO coronographic images. Differential imaging techniques are nowadays a must-have for any high-contrast instruments to address that the residual speckles limitation. Various strategies have been adopted including: (i) reference differential imaging (RDI) using a reference single star observed before/after the science target, (ii) angular differential imaging (ADI) exploiting the field and pupil rotation of alt-az Telescope, (iii) spectral differential imaging making use of the different spectral signatures between the planet and the star completed by molecular mapping techniques combining high-contrast and spectral cross-correlation methods, (iv) finally, polarimetric differential imaging well adapted to image the scattered polarized light of planet-forming and debris disks (or the reflected polarized light of exoplanets). Please refer to [7] for more details on the limitation and mitigation for the detection and characterization of the exoplanetary signal with planet imagers.

### 1.3. *Young, nearby stars identification*

For more than two decades now, planet imagers have been mainly targeting young (≤500 Myr), nearby (≤100 pc) stars to search for giant exoplanets and disks. This is motivated by the fact that young exoplanets are hotter, brighter and easier to detect than older counterparts, and that the system proximity allows the exploration of the close physical surrounding. The identification of these comoving groups of young stars started from an anomaly observed in the 80's. The TW Hya T Tauri star was discovered isolated from any dark cloud nor birth place regions [8]. A few years later, IRAS excesses combined with H$\alpha$ and Lithium depletion measurements enabled the identification a handful of young stars in the vicinity of TW Hya. Their origin, runaway stars from some cloud or formed in situ from a low-mass cloud, their age and their distance remained unclear for years. X-ray data finally confirmed that they were members of the first identified young, nearby association, i.e. the nearest known region of recent star formation, the so-called TW Hydrae association [9] (TWA). Over the next decade, the number of TWA members drastically increased to more than 30 members, with age and mean distance estimates converging towards 10 Myr and 50 pc, respectively. Immediately after the confirmation of the existence of the TWA association, several research groups became specifically interested in identifying new young, nearby associations in the vicinity of the Sun [10, 11].

The key criteria of identification relies on two main pillars studying: (i) the galactic position and motion of a star to derive its age. The galactic kinematics studies have become increasingly sophisticated using for instance Bayesian analysis to derive membership probability of a star and its age considering combined photometric and kinematics catalogs, (ii) the spectro-photometric stellar characteritics in relation with well calibrated age-sensitive indicators. Various youth diagnostics can be used depending on the stellar age and spectral type like color-magnitude diagram, rotation, Lithium depletion, H$\alpha$ emission, Ca H&K, Coronal X-rays and chromospheric

---

[2]The IWA is universally defined as the 50% off-axis throughput point of a coronographic system.

UV activity, or the presence of IR excess to inter-compare and refine stellar ages. Nowadays, more than 500 young, nearby (<100 pc) stars have been identified in the Sun vicinity, mostly F, G, K and M dwarfs. Systematic studies, including the recent *Gaia* results, are now pushing the horizon down to very low-masses with the discovery of a large population of M-, L- and even T-type members [12]. These young, nearby stars are gathered in several groups (TW Hydrae, $\beta$ Pictoris, Tucana-Horologium, $\eta$ Cha, AB Dor, Columba, Carinae...), sharing common kinematics, photometric and spectroscopic properties. This horizon has been extended to members of intermediate-age associations or star-forming regions located at larger distance like the so-called Scorpius Centaurus complex composed of thousands of stars. With typical contrast of 10–15 magnitudes for separations beyond 200 mas (10–20 au for a star at 50 pc) achieved with the second generation of planet imagers, young giant planets down to 1–2 Jupiter masses are typically detectable. Therefore, without surprise, a significant amount of telescope time was dedicated to deep imaging surveys of young, nearby stars to search for exoplanets, brown companions and disks leading to the first direct imaging discoveries of planetary mass companions in the early-2000's.

## 2. Results

Today, deep direct imaging campaigns have surveyed several hundred members of young, nearby associations (TW Hydrae, $\beta$ Pictoris, Tucana-Horologium, Columba, Argus,...) and star-forming regions (Lupus, Taurus, Ophiuchus, Chamaeleon, and Sco-Cen). Various motivations were followed for the target selection of these surveys: (i) complete census of given associations, (ii) selection of young, intermediate-mass stars, (iii) or very low-mass stars, or (iv) application of figure of merit considering detection rate with toy models of planet population. The near-infrared wavelengths have been used intensively, particularly with SPHERE and GPI lately. They are a good compromise between low-background noise, angular resolution and good Strehl correction. However, thermal (3–5 µm) imaging has been very competitive in terms of detection performances as the planet–star contrast and the Strehl correction are more favorable at those wavelengths despite an increased thermal background. The first clues of the presence of a giant planet directly imaged is revealed by the presence of a faint point-source, presenting cool atmosphere features as described below, in the close vicinity of a young star. Very red near-IR colors or specific spectral features such as a peaked *H*-band spectrum are indicative of young L dwarfs atmospheres. Methane absorption indicates a probable young and cool T dwarf. These diagnostics enable the observers to rapidly estimate the predicted mass, effective temperature and physical separation of the candidate using evolutionary model predictions. Robust confirmation then come with follow-up observations at consecutive epochs to verify that the candidate is actually co-moving with the central star considering its parallactic and proper motion. Orbital motion can be resolved to unambiguously confirm that both objects are physically bound. Independently of the technique used, these surveys have shown that imaging exoplanets remains a challenging task as the typical occurrence rate of giant planets, more massive than a few Jupiter masses and beyond 10 au, in nearby systems is a few percent at most [13]. It therefore requires the observation of large samples (typically 500 stars for SPHERE and GPI exoplanet surveys) with a significant amount of observing time, strongly contaminated by background stars representing the vast majority of the point-source detection. They however led to breakthrough discoveries over the past decades (see illustrative gallery in Figure 3) and a better understanding of the processes of planetary formation and evolution, the physics of young giant planets, and of their demographics as described below.

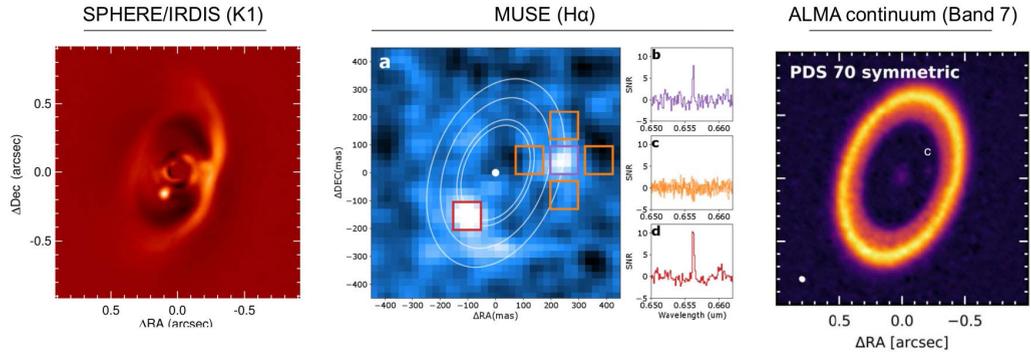

**Figure 3.** Left, SPHERE combined $K1K2$ image of PDS 70 b inside the transition disk surrounding PDS 70 [14]. Middle, MUSE observations of PDS 70 b and c [15]. Right, ALMA observations of the disk and the CPD around PDS 70 c [16].

## 2.1. *Planet-forming disks, protoplanets & bulk properties*

In addition to the presence of planet in formation, planet imagers offer us unprecedented information about the morphology of young planet-forming disks during or right after the planetary formation phase. Various substructures like large spiral arms, multiple rings, shadows or cavities can be spatially resolved and in some cases characterized spectroscopically or polarimetrically [17]. Several sharp, Kuiper-like belts were also imaged [18, 19]. These first studies paved the way towards the first unambiguous discovery of two recently formed young planets within a transition disk surrounding the young Sun-like star PDS 70 ([20]; see Figure 4). In synergy with ALMA and MUSE at VLT, SPHERE observations offer a fabulous way to directly explore the planetary formation and evolution, the atmospheric and orbital properties of young Jupiters [14], the physics of planetary accretion [15, 21], the presence and properties of circumplanetary disks [16], and the global system architecture to study the gas and dust dynamics [22]. These discoveries allow us to simultaneously connect the giant planet physical and orbital properties with the disk morphology, and potentially the inner regions where terrestrial planets might form (see [23]).

Today, one key limitation of direct imaging is that we do not directly measure the mass of these young exoplanets, but their photometry, luminosity and spectral energy distribution. Consequently, we have to rely on evolutionary model predictions that are not well calibrated at young ages to convert luminosity to mass. In addition to the system age uncertainty, the predictions highly depend on the formation mechanisms and the gas-accretion phase that will form the exoplanetary atmosphere. The way the accretion shock will behave (sub or super-critical) on the surface of the young accreting proto-planets during the phase of gas runaway accretion will drive the planet initial entropy or internal energy, hence its initial physical properties (luminosity, effective temperature, surface gravity and radius) and their evolution with time [36]. These different physical states are described by the so-called hot-start (sub-critical shock), cold-start (super-critical shock), and warm-start (intermediate cases) models. They predict luminosities that are spread over several order of magnitudes for young, massive giant planets. Over time, these young giant planets will then cool down and shrink as they are not massive enough to sustain Hydrogen nuclear fusion. Consequently, a lot of open questions remain to be answered about the formation and the evolution of giant planets and of their atmospheres from the theoretical point of view. The observations of proto-planets in their birth environment and the direct measurement of emission lines have just started, and allow to explore the planetary accretion processes and shocks to constrain the way the gas is accreted on young forming planets [37]. The synergy of techniques, combining direct imaging and/or interferometric observations with radial velocity and/or astrometry with *Gaia* is particularly powerful to constrain the Mass-Luminosity relation of young Jupiters as recently done for β Pic b & c [38] or HR 8799 e [39]. The future *Gaia* Data Release 4 (2025), that will soon deliver fully calibrated astrometric time series, is in that sense very promising as it will pinpoint the presence of young giant planets between 3–10 au orbiting young, nearby stars in the vicinity of the Sun, offering prime follow-up targets for planet imagers to explore the diversity of the bulk properties of young giant planets in connection with their environment.

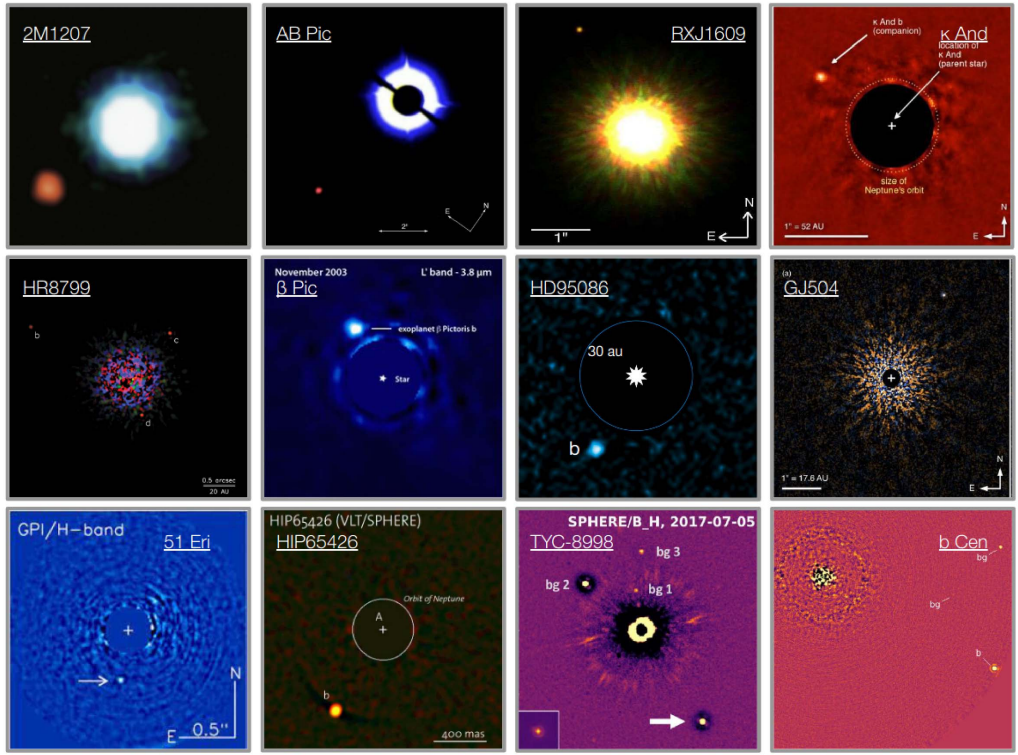

**Figure 4.** Emblematic direct imaging discoveries (not exhaustive): 2M1207 & AB Pic [24,25], RXJ 1609 [26], κ And [27], HR 8799 [28], β Pic [29], HD 95086 b [30], GJ 504 [31], 51 Eri b [32], HIP 65426 b [33], TYC 8998-760-1 b [34], and b Cen b [35].

## 2.2. *Exoplanetary atmospheres*

Since 2014, the use of integral field spectrographs with SPHERE and GPI enabled to acquire for the first time low-resolution ($R_\lambda \sim 50$) emission spectra of tens of young giant planets exhibiting broad features due to unresolved molecular absorptions ($H_2O$, $CH_4$, VO, FeH, CO). These spectra can be compared to predictions of atmospheric models to infer first-order information on the bulk properties of the atmosphere, mainly the effective temperature, the surface gravity (pressure) and the properties of clouds. The current atmospheric models used by the community are mainly 1D-models involving at least hydrodynamics, radiative and convective energy transport, and gas-phase chemistry. They solve the pressure–temperature structure in radiative-convective equilibrium and determine the chemical species that are supposed to form. Sophisticated models of clouds of different composition (silicates, sulfites…) have been considered for more than a decade (refer to [40] for more details). The analysis of these low-resolution spectra of young giant exoplanets showed that, as for young brown dwarfs, young exoplanets of a given spectral type with lower surface gravity can be up to 200–500 K cooler than their older counterparts [41, 42]. This discontinuity is particularly noticeable at the M–L transition, and also evidenced by a lack of methane at the L/T transition as seen for HR 8799 b [43]. These results confirm the early photometric characterization that showed the peculiar properties of young L/T type planets like 2M1207 b, HR 8799 bcde, HD 95086 b or HD 206893 b owing to the low-surface gravity conditions in their atmospheres, leading to an enhanced production of clouds probably composed of sub-µm dust grains made of iron and silicate ([44]; see Figure 5). With cooler temperatures than typically 1000 K, they start dissipating and fragmenting, modifying the cloud vertical and spatial distribution as suggested by recent variability studies showing high-amplitude photometric variations for several of these young, planetary-mass companions [45]. The presence of circumplanetary disks, as directly evidenced for PDS 70 c by ALMA sub-mm observations, might also affect the spectral energy distribution of the youngest planets mixing contribution in infrared from the photosphere and the (spatially unresolved) circumplanetary disk [16, 46].

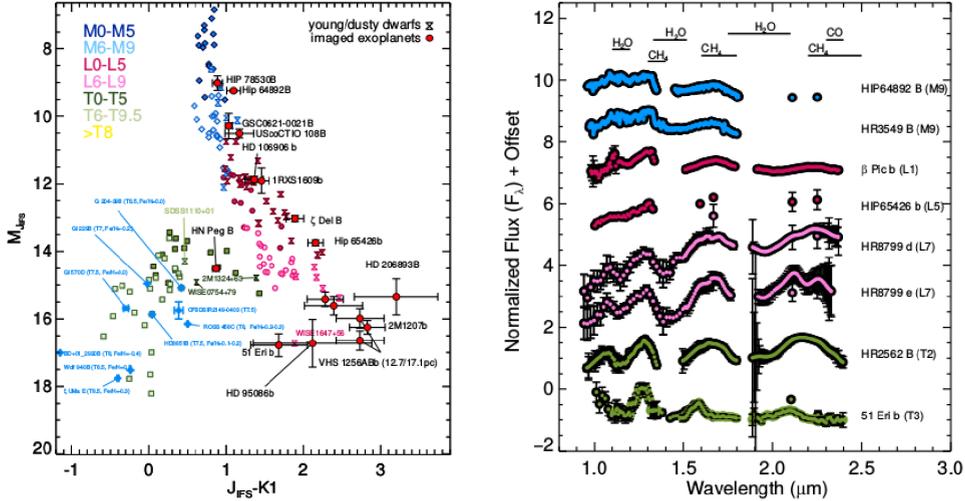

**Figure 5.** Left: color-magnitude diagrams comparing the young imaged exoplanets with field M, L, and T dwarfs with known parallax measurements. Right: Sequence of exoplanet's spectra characterized in the course of the SHINE and GPIES surveys. Adapted Figures from [47].

With the increase of spectral resolution ($R_\lambda > 1000$), the line blending of molecular absorptions start to be resolved yielding unprecedented accuracy constraints on surface gravity (pressure), chemical abundances and the complex molecular chemistry of exoplanetary atmospheres (see Figure 6). Atmospheric model degeneracy can be better suppressed and missing opacities identified. Note that the combination of spectral diversity with higher spectra resolution and high-contrast imaging is particularly interesting to boost the detection performances of young exoplanets using the so-called *molecular mapping* technique [48]. Regarding the characterization part, high-contrast spectroscopic observations at medium resolution with AO-fed spectrographs like OSIRIS at Keck, SINFONI at VLT, led to the determination of molecular abundances such as water, carbon-monoxide or methane for several young planets like HR 8799 b and e, $\beta$ Pic b, HIP 65426 b [49–52]. More recently, even the $^{12}CO/^{13}CO$ isotopologue ratio, connected to the carbon-ice fractionation process, was measured for the first time in the atmosphere of the young exoplanet TYC-998-760-1 b suggesting a formation beyond the CO snowline [53].

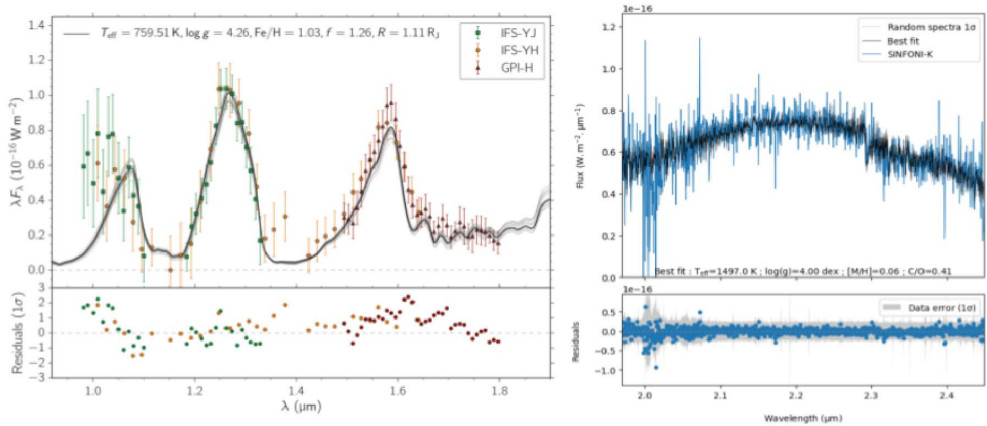

**Figure 6.** Left, Spectral energy distribution for 51 Eri b, a young, T3-type, 2–3 $M_{Jup}$ giant planet orbiting at 11.1 au from 51 Eri (F8V), constructed from SPHERE and GPI observations [55, 56]. Predictions from the petitCODE cloudy model are reported with the best fitting solution of the bulk properties (and 32 randomly drawn samples from the posterior probability distribution in gray). Right, K-band SINFONI spectrum of the warm, dusty planet HIP 65426 b, with the best ExoREM fit model found.

Accessing the planet atmospheric composition (C/O, D/H, N/O, N/C and isotopologue ratios) and metallicity is a path forward the exploration of the formation mechanisms, the birth location and the potential migration/dynamical history of the exoplanets. The observed atmospheric composition is however resulting from various physical complex processes. They are connected to the disk composition, physical and thermal structure, chemistry and evolution that will define the building blocks composition of the giant planet, the planetary formation processes that will affect the solid-gas accretion phase, the vertical mixing/diffusion between the planet's bulk and atmosphere, or simply the dynamical evolution over time, which overall will make any direct interpretation speculative at this stage [54]. An interesting comparison point is provided by the population of young free floating brown dwarfs (or exoplanetary analogs), that share common bulk properties in terms of mass, temperature, surface gravity and age. Less challenging to observe and characterize, they offer a rich and complementary test bench for atmospheric models. They offer the opportunity to explore the chemical differences with the population of young giant planets that could trace different processes of formation.

## 2.3. *Orbit, architecture and dynamical evolution*

All young giant planets discovered in direct imaging up to now orbit beyond 5 au, revolving around their stellar host with typical periods longer than 10 years. After their discoveries, intense monitoring and data mining enabled to rapidly derive orbital solutions of the most favorable cases. For $\beta$ Pic b, since its first image in November 2003, almost 20 years of NaCo and SPHERE, GPI, and more recently VLTI/GRAVITY astrometric measurements have been carried out covering almost 100% of its orbit [38, 57]. For HR 8799, the longest baseline has been achieved for the b planet thanks to the re-analysis of archived *HST/NICMOS* data from 1998 offering a time coverage of almost 25 years now combining NIRC2, GPI, SPHERE and more recently GRAVITY measurements (see Figure 7). Up to 25% of the orbit for the three closest planets (cde) has been covered so far [58]. For the remaining imaged planets, the orbital motion is generally resolved but covers a small fraction ($\lesssim$10%) of the orbit, sometimes finding hints of orbital curvatures as observed for 51 Eri b, GJ 504 b or HD 95086 b, and therefore limiting the precise determination of their orbital parameters. Accurate astrometric monitoring in direct imaging is a meticulous task due to the relatively small field of view of these instruments. The second generation of planet imagers like GPI, SCExAO and SPHERE now regularly achieves astrometric precision down to 1–2 mas with adequate calibration strategies, compared to the 10–20 mas precision achieved with the first generation of planet imagers [59]. The recent advent of the dual-field astrometric mode of GRAVITY at VLTI is now out performing all planet imagers in terms of astrometric precision by providing measurements at the 100 to 10 µas level (see [3]). Direct imaging (and interferometric) orbital monitoring are currently completed by radial velocity and *Gaia* astrometric studies whenever possible [39]. They can be combined with direct measurement of the planet's radial velocity using medium to high-resolution spectroscopy [60] to study the planetary orbits in the three dimensions, and even explore their obliquities and alignment with the stellar spin [61, 62].

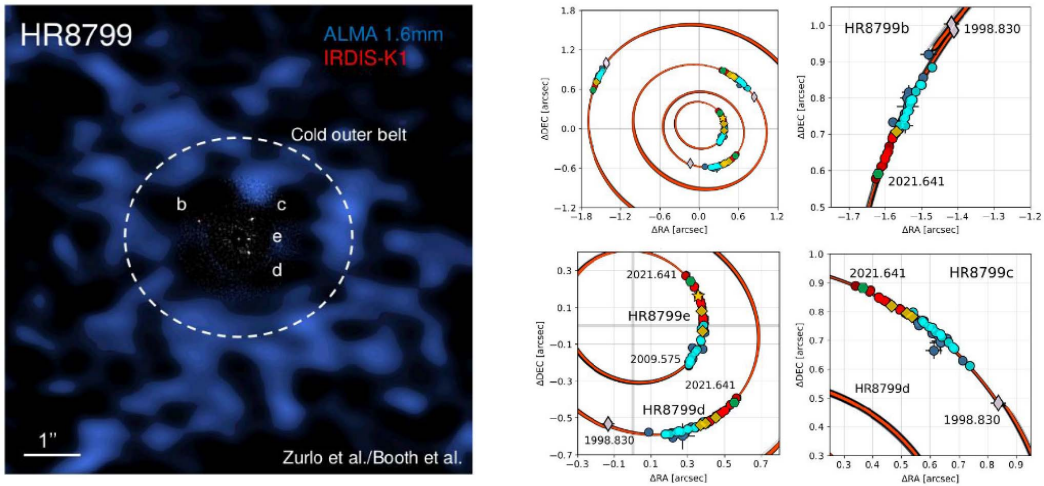

**Figure 7.** Left, Composite ALMA/SPHERE image of the HR 8799 system. ALMA at 1.6 mm reveals a cold component of dust, Kuiper belt-like, beyond 100 au, while the four giant planets are distinguishable in the central part thank to SPHERE/IRDIS at K1-band in infrared. Right, Motion of the HR 8799 bcde planets monitors by SPHERE (red-filled circles and yellow hexagons), LUCI at LBT (green hexagons), NIRC2 at Keck (light-blue filled circles and dark-blue circles), GPI (Diamonds) and a star symbol is for the most accurate GRAVITY point. The best-fitting solutions to the four-planet model is reported in red. Figures adapted from [63, 64].

The precise determination of the orbital properties of young giant planets is an important step towards the study of the structure and dynamical stability of planetary systems, including the planet–planet and planet–disk interactions. This is directly connected to our understanding of the formation of giant planets, but also to the formation of terrestrial planets as the giant planets (when present) will shape the global system architecture and dynamical evolution. Early studies, in the case of $\beta$ Pic b, evidenced for the first time a clear planet–disk interaction showing that the planet is responsible for the inner warp deformation of the debris disk around $\beta$ Pic, forcing the planetesimals to precess on a 2–5° misaligned orbit with respect to the main disk [57, 65]. The planet location and low, but non-null eccentricity, is also consistent with the secular dynamical pertubations at the origin of the exocometary activity monitored for more than 30 years for $\beta$ Pic [66]. Signs of interactions are indirectly seen in the substructures of debris disks (waves, eccentricity, excess of exozodiacal dust), or at even younger ages (cavities, vortices, spirals, ... ). However, there are only a few cases for which the disk sub-structures and the planets are both directly imaged and can be unambiguously connected, and studied. The three emblematic systems HR 8799, HD 95086 and PDS 70 are particularly interesting in that sense. The giant planets have all low-eccentricities, and orbit between two coplanar dusty belts, similarly to the giant planets of our Solar System orbiting between the asteroids and Kuiper belts [67]. For the multiple-planetary systems HR 8799 and PDS 70, the giant planets orbit in a configuration close to mean motion resonances that do favor dynamical stability [68, 69]. Interestingly, a correlation between the formation of terrestrial planets in the inner region of the planetary systems, the presence of cold debris dust and the existence of stable low-eccentricity giant planets has been predicted based on a comprehensive set of dynamical simulations [70, 71]. Systems with high-eccentricity giant planets are on the contrary not expected to host debris disks and inner terrestrial planets, or could indicate a stellar-like architecture and formation pathway [72]. Consequently, finding more of these young, stable Solar system analogs, understanding how common they are, is fundamental to understand if the architecture of our Solar system is a rare product (or not) of planetary formation, and if it is an essential condition for the formation of life.

## 2.4. *Demographics*

Considering the very large number of young, nearby stars observed and the dozens of imaged exoplanets discovered so far, the rate of discovery in direct imaging remains relatively low (typically ~5–10%), but relatively similar to the rate of inner giant planets discovered in radial velocity or transit. A robust statistical analysis including the survey completeness and the confirmed exoplanets can be exploited to derive the occurrence of the young giant planet population beyond typically 10 au. For that purpose, early studies have developed statistical analysis tools to exploit the complete deep imaging performances and derive the detection probabilities of their surveys by simulating synthetic planet population described by various sets of parametric distributions (mass, eccentricity, semi-major axes) [73–75]. They show that the first generation of surveys were mostly sensitive to giant planets more massive than 5 $M_{Jup}$ for semi-major axis between typically 30 to 300 au. They typically derive overall occurrence rate for planetary systems hosting at least on 5–13 $M_{Jup}$ exoplanets of less than 4–6% (assuming hot-start evolutionary models) around FGK-type stars. Hints of a dependence of this occurrence rate with the stellar host mass, analogous to the well-established correlation observed at small separation with radial velocity surveys, was suggested by surveys specifically targeting intermediate-mass stars and low-mass stars [30, 76, 77].

With the improved sensitivity in mass and semi-major axis, GPI and SPHERE enable a significant step forward by providing meaningful sensitivity to planetary-mass companions and brown dwarfs down to 10 au, getting close to the snow line region where the bulk of the giant planet population formed by core accretion is expected. Ref. [78] identify for instance a break in the radial velocity giant planet distribution between 2 and 3 au after which the occurrence rate decreases with distance from the star, but which must be confirmed with direct imaging and future *Gaia* astrometric results to get a global picture of the population of giant planets at all separation. With SPHERE and GPI, it finally becomes realistic to compare direct imaging observations to predictions of both core accretion models coupled to dynamical planet scattering evolution, and gravitational instability models [13, 79]. The first statistical analysis on the properties of the population of sub-stellar companions at wide orbital separation based on a sub-set of 150 stars for SPHERE gives a frequency of systems with at least one companion with a typical mass in the range $M_p$ = 5–75 $M_{Jup}$ and a semi-major axis in the range $a$ = 10–300 au, to be $5.7^{+3.8}_{-2.8}$% for FGK stars (see Figure 8). Qualitatively, the contribution of the core accretion part of the model seems enhanced over the gravitational instability part, which means that core accretion contributes a higher fraction of the full companion population required to explain the data. The use of parametric models of planet and brown dwarf population also confirms (i) that the frequency of substellar companions is significantly higher around BA stars than around FGK and M stars, (ii) and that the planet-like formation and brown dwarf binaries-like formation mechanisms overlap, but dominate over different ranges of planet-to-star mass ratios, $q$, which is a strong constraint for formation models.

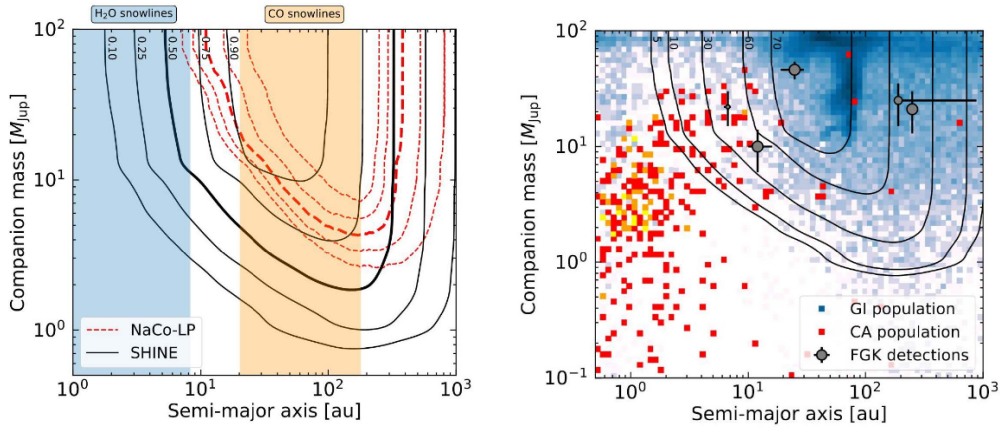

**Figure 8.** Left, Comparison of the sensitivities of the NaCo and SPHERE exoplanet surveys, based on the average probability of detecting a companion as a function of its mass and semimajor axis. The range of semimajor axes spanning the $H_2O$ and CO snow lines for the stars in the sample are overplotted. Right, Comparison of the depth of search of the SPHERE exoplanet survey with the population synthesis models based on the CA and GI formation scenarios presented. The CA companions are represented with shades of red (low density of companions) to yellow (high density of companions), and the GI companions are represented with shades of white (low density of companions) to blue (high density of companions). Figure from [13].

## 3. Perspectives

In less than two decades, exoplanetary science has become a major field of fundamental research in physics, recently rewarded with a first Nobel Prize. This is well illustrated by the pace of new discoveries and breakthrough results incessantly impacting the community and the general public. Two decades ago, the only planets we knew were the ones of our Solar System. Today, thousands of exoplanets have been discovered since the 51 Peg discovery [80]. These new worlds reveal an incredible diversity (Hot Jupiters, irradiated and evaporating planets, misaligned planets with stellar spin, planets in binaries, terrestrial planets in habitable zone, discovery of Mars-size planet, circumplanetary disks possibly hosting exomoons in formation...), driving current theories of planetary formation and evolution. On the timescale of a human Life, we may well be witnessing the first discovery of an exoplanet and the first detection of bio-signatures in the atmosphere of a nearby exo-Earth!

### 3.1. *A Golden Era for Exoplanets*

Exoplanetary science is today a major workhorse for the astronomical community, driving the scientific and instrumental roadmap, in Europe, for the development of future space missions and ground-based telescopes and instruments for the decades to come. Covering various domains of expertise from theory, simulation, instrumentation, lab experiments to observation, exoplanetary science naturally diversified in synergy with the study of the Solar System and the stellar physics, but also transversally connecting with other science fields like biology, chemistry, geophysics, climatology, photonics, signal processing, artificial intelligence . . . . Despite this success, we do not have at this stage a global understanding of the formation of giant and terrestrial planets to explain the diversity of the new worlds routinely observed, of the place our Solar System in this exoplanetary bestiary, and of the seminal conditions required for the emergence of life. Many fundamental questions remain to be answered concerning: (i) the existence of one or several mechanisms to form giant planets, (ii) the physics of accretion driving the physical properties of young Jupiters and their evolution over time, (iii) the impact of dynamical evolution in crafting planetary system architectures with favorable conditions for life, or (iv) the (non-)singularity of Earth within all habitable worlds. In that perspective, the upcoming decade is rich in terms of space missions and ground-based instrumentation that will attempt to answer these fundamental questions (see Figure 9).

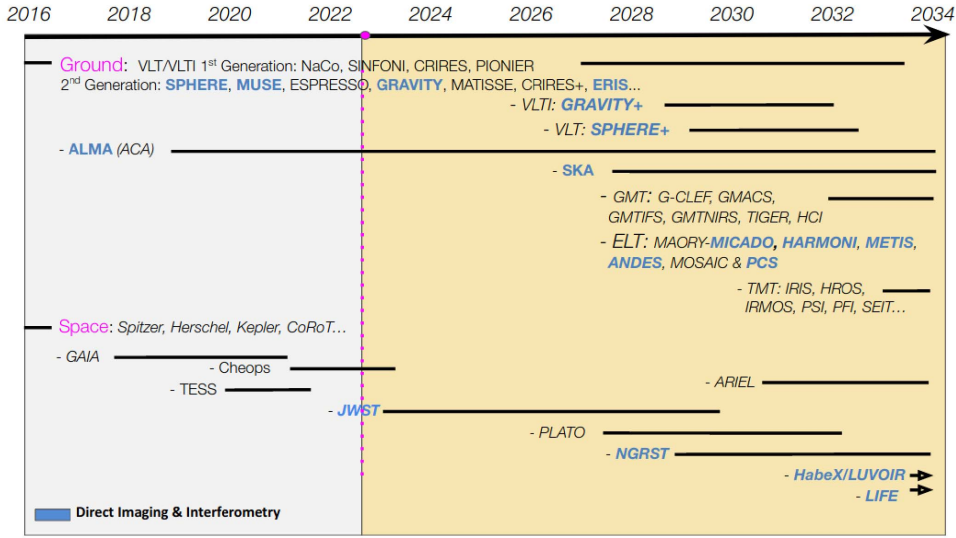

**Figure 9.** Timeline of current and future space and ground-based projects with a strong invovlement fo the European community and dedicated to planetary formation and exoplanetary science.

Following the second generation of planet imagers (GPI, SPHERE & SCExAO), new instruments are coming on line to push the exploitation of the spectral diversity in connection with the spatial one. A new set of medium resolution integral-field spectrographs and upgrades in visible and/or infrared, like MagAO-X at Magellan (early-2022), ERIS at VLT (mid-2022), SHARKS-VIS and NIR at LBT (late-2022), MAVIS at VLT (2028), will start offering new capabilities to image and characterize young exoplanets in accretion and/or in molecular mapping with upgraded xAO facilities. They should provide new insights on the connection of disk sub-structures with young protoplanets in formation, but also on the chemical abundances of imaged exoplanets. The coupling of (x)AO system with high-resolution ($R_\lambda$ = 30,000–100,000) fiber-fed spectrographs has also started with CRIRES(+) (soon HiRISE at VLT coupling SPHERE and CRIRES+, and foreseen for late-2023) and KPIC at Keck (since 2020). They offer unprecedented opportunities to systematically measure the radial and rotational velocities of the imaged planets to determine their or-bit in three dimensions, their true obliquity (when coupled to photometric follow-up), and to open new horizons to map the chemical composition, and the general atmosphere circulation of young Jupiters. To push the survey capabilities of SPHERE and GPI to access the bulk of the young giant planet population down to the snow line (3–10 au), upgrades of their xAO systems and instruments have been proposed [81, 82]. Both upgrades aim also at observing a larger number of fainter, redder stars in the youngest star-forming regions, building on the synergy with ALMA to characterize the architectures and properties of young planetary systems in formation. They want to improve their characterization capabilities by either increasing the spectral wavelength coverage or the spectral resolution of their integral-field spectrographs. In parallel, on-going developments at VLTI with GRAVITY+ (2026) and the ASGARD visiting instruments will expand the fringe tracking and sensitivity performances, the wavelength coverage (JHKL-bands) and spectroscopic modes, and utlimately the interferometric contrast implementing nulling techniques (see [3]).

For the decade to come, this set of new planet spectro-imagers (and interferometers) will work in concert with complementary facilities connected to Exoplanetary science. At sub-mm and centimetric wavelengths, ALMA (and potential upgrades) will pursue the characterization of the cold dusty and gaseous component of young planet-forming and debris disks with an exquisite spatial resolution similar to the one achieved by current planet imagers in infrared (down to 10 mas), and will be soon completed by the Square-Kilometer Array (SKA) starting operation in the late-2020. The new generation of high-resolution spectrographs dedicated to radial velocity studies in visible ESPRESSO, EXPRES, HARPS3, and infrared CARMENES, SPIROU, NIRPS, will extend the current horizon to the population of terrestrial planets around solar and low-mass stars (including young stars) and dedicated strategies to tackle the problem of stellar activity. In space, *Gaia* will achieve a final astrometric precision of 10 µas in the context of a systematic survey of a billion of stars and therefore discover thousands of new planetary systems. The *Gaia* Data Release 4 (in 2025) will give us a complete census of the young and old giant planet population between typically 2 and 5 au for stars closer than 200 pc. Dedicated *Gaia* follow-up in direct imaging have already started based on the proper motion anomaly measured between *Hipparcos* and *Gaia* indicating the potential presence of substellar perturbers [83]. Going beyond the the *Corot*, *Kepler*, *TESS* and *CHEOPS* missions, *PLATO*, foreseen for 2026, will extend our knowledge on the content of terrestrial planets at longer periods, up to several years, around relatively bright, nearby stars. Within 10 years, the era of large-scale systematics surveys will decay thanks to a complete census of exoplanetary systems within 100–200 pc from the Sun. A new Era fully dedicated to the characterization of known systems will rise. Already initiated with *Hubble*, *Spitzer* and the first and second generation of planet imagers and spectrographs, the characterization of the physics of giant and terrestrial planets, has been done today for about 30 exoplanets from super-Jupiters, Hot Neptunes to super-Earths. The presence of water, carbon monoxide and methane molecules, of haze revealed by Rayleigh scattering, observation of day-night temperature gradients, constraints on vertical atmospheric structure and atmospheric escape have been evidenced in the past decade. This will intensify with the James Webb Space Telescope (*JWST*), that started operation in mid-2022. *JWST* will start revolutionize transit and secondary eclipse spectroscopy by exploring a broad diversity of exoplanetary atmospheres over a wide range of mass, temperature, and composition. Despite a design not dedicated to high-contrast imaging, the *JWST* instruments NIRCam and MIRI are equipped with coronographs. NIRSpec and MIRI can also exploit the spectral diversity provided by their IFS ($R_\lambda$ = 2700 and 1500–4000, respectively) modes for the detection and characterization of exoplanets [84]. At separations larger than 0.6–1.0 as, *JWST* will be unique for the exploration of the outer part of young, nearby planetary systems, to detect young Saturn-analogs beyond typically 30–50 au down to Saturn-mass thanks to its unprecedented sensitivity. Finally, *ARIEL* (launched in 2029) will be fully devoted to the systematic characterization of exoplanetary atmospheres by surveying a diverse sample of about 1000 extrasolar planets, simultaneously in visible and infrared wavelengths, and will be almost contemporaneous to the first Lights of the extremely large telescopes (ELTs).

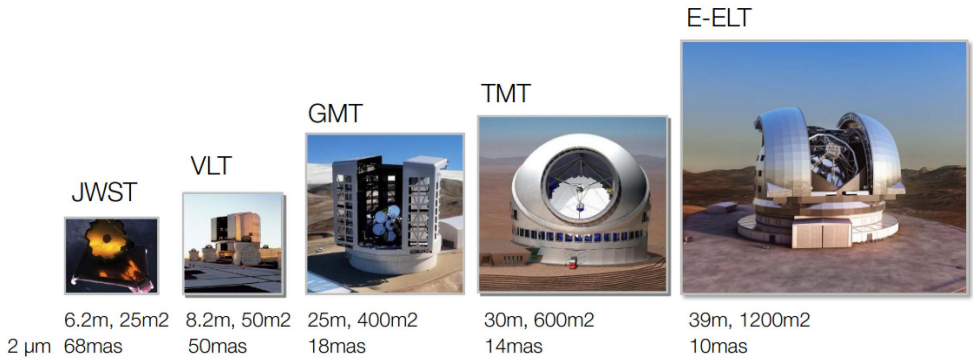

**Figure 10.** The VLT 10-m class telescope, together with the James Webb Space Telescope and the new generation of extremely large telescope, the Giant Magellan Telescope, the Thirty Meters Telecope and the European Extremely Large Telescope (hereafter E-ELT, [71]). Telescope diameters, collecting areas and spatial resolution at 2 µm are reported.

3.2. *The Age of Extremely Large Telescopes*

With first light foreseen for 2028 and after, the ELTs, the Giant Magellan Telescope (GMT, [85]), the Thirty Meters Telecope (TMT, [86]) and the European Extremely Large Telescope (hereafter E-ELT, [87]), will offer unique capabilities in terms of sensitivity, spatial resolution and instrumental versatility (see Figure 10). They represent some of the most challenging projects in modern astronomy with the realization of 30 to 40 m-class Adaptive Telescopes, i.e. with AO included in the telescope design, conceived for visible and infrared wavelengths and equipped with segmented primary mirrors (7 segments of 8.4 diameter for GMT, 492 segments of 1.44 m for TMT and 800 segments of 1.45 m for E-ELT). They will rely on a diversity of dedicated instruments offering various observing modes: from integrated field spectroscopy, high-precision astrometry, simultaneous multi-object spectroscopy of hundred of targets, high resolution spectroscopy or high contrast imaging over a broad range of wavelengths (from 0.33 to 19.0 µm) together with a high operation efficiency. Various flavors of AO systems will be used to ultimately provide the community with diffraction limited images down to ∼10 mas in *K*-band and/or to exploit the full patrol field of view of several arcminutes offered by these telescopes.

The increased sensitivity and exquisite spatial resolution of the ELTs will offer an incredible opportunity to directly scrutinize and characterize the stellar environment at unprecedented physical scales, from the inner regions of proto-planetary disks, the chemistry of planet-forming zones, to the characterization of recently formed giant planets in their birth environment. Figure 11 illustrates the synergy between current 10 m-class telescope instrumentation, ALMA and the future instrumentation of the ELTs to characterize the spatial, temporal and chemical evolution of young planet-forming disks. The spatial distribution of dust and gas, including asymmetries and over densities in planet-forming zones, will be explored at a 1–10 mas scale (10–100 µas scale in spectro-astrometry). The spectral information with a resolving power of 100 000 will enable to directly map the gas dynamics with a velocity precision of 3 km·s$^{-1}$ (for instance CO with METIS at 4.7 µm). Keplerian rotation will be distinguished from wind, accretion or Jet components. Deviation from Keplerian might also help distinguishing the presence of hidden planets. The inner disk chemistry, the disk atmospheres, the physical transport of volatile ices either vertically or radially, and the importance of non-thermal excitation processes in the planet-forming regions will be also explored. A direct view of the distribution and the dynamics of water, playing a key role for the planetesimals formation and the disk cooling, will be possible. The organics content like $CH_4$, $C_2H_2$, HCN in the planet-forming regions and the prebiotic chemistry will be studied. Ultimately, the study of isotopic fractionation should enable to probe the chemical and physical conditions in the planet-forming disks to improve our understanding of the transfer processes of water on terrestrial planets.

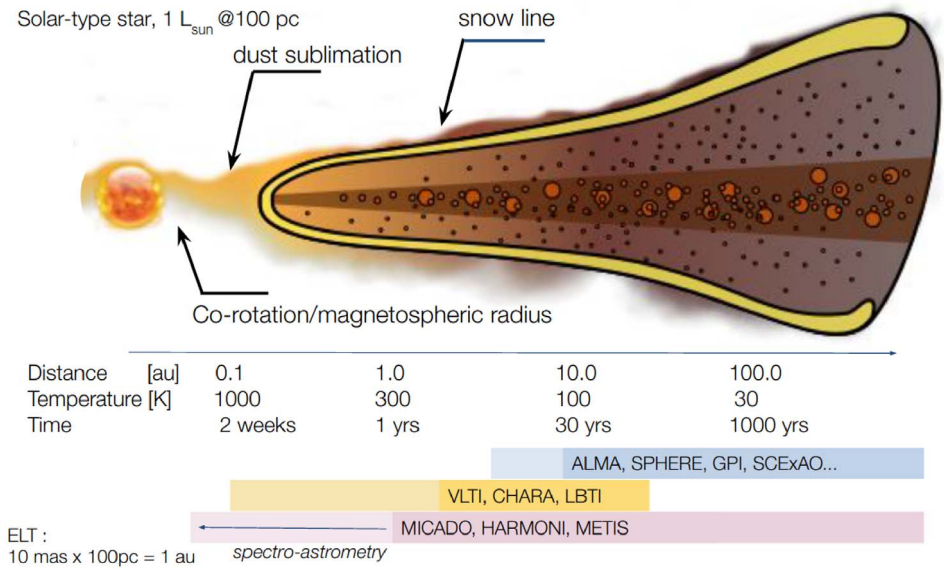

**Figure 11.** Sketch of the parameter space explored with ELTs and other facilities. Direct spectral imaging of the main planet-forming regions will be achievable with the ELTs. For a typical young star at 100 pc, a 10 mas spatial resolution corresponds to physical separations of 1 au, i.e. to the exploration of the warm gas and dust spatial distribution down to the snow line, the disk evolution and dissipation to ultimately determine the initial conditions of planetary formation.

By 2028, the ELTs will arrive at a propitious time where thousands of new planetary systems, in the Solar vicinity, will have been discovered and characterized by a large set of instruments or space missions, therefore covering a broad range of the parameter space in terms of physical properties (planetary masses, semi-major axis, radii, density, luminosity, atmospheric composition) and stellar host properties (age, mass, binarity, composition . . . ). The ELTs will therefore be mainly used for the detailed exploration of known planetary systems, to search for additional planets or characterize the most interesting ones. First light instruments, like MORFEO/MICADO and HARMONI at E-ELT, will be multi-purpose instruments, not designed for high-contrast, and operating with single-conjugated AO modules offering limited performances in term of Strehl (60–80% at K-band), but already offering the perspective to exploit the spatial and spectral diversity of their spectro-imagers. They will reach typical contrasts (5 $\sigma$) of $10^{-6}$ at angular separations of 50–100 mas at H and K-bands [88, 89]. This will enable to completely bridge the gap with indirect techniques like radial velocity and astrometry for a systematic exploration of the population of young, self-emitting giant planet at all separation, to tightly constrain the Mass-Luminosity relation of young Jupiters, and to boost the exploration of their atmosphere by getting high-fidelity spectra at medium to high-spectral resolution. Given the ELT sensitivity, Doppler imaging techniques will be applied to directly map the spatial distribution of diverse molecular lines (CO, $CO_2$, $H_2O$, $CH_4$ and possibly $NH_3$), and explore the general circulation of their atmospheres (see Figure 12).

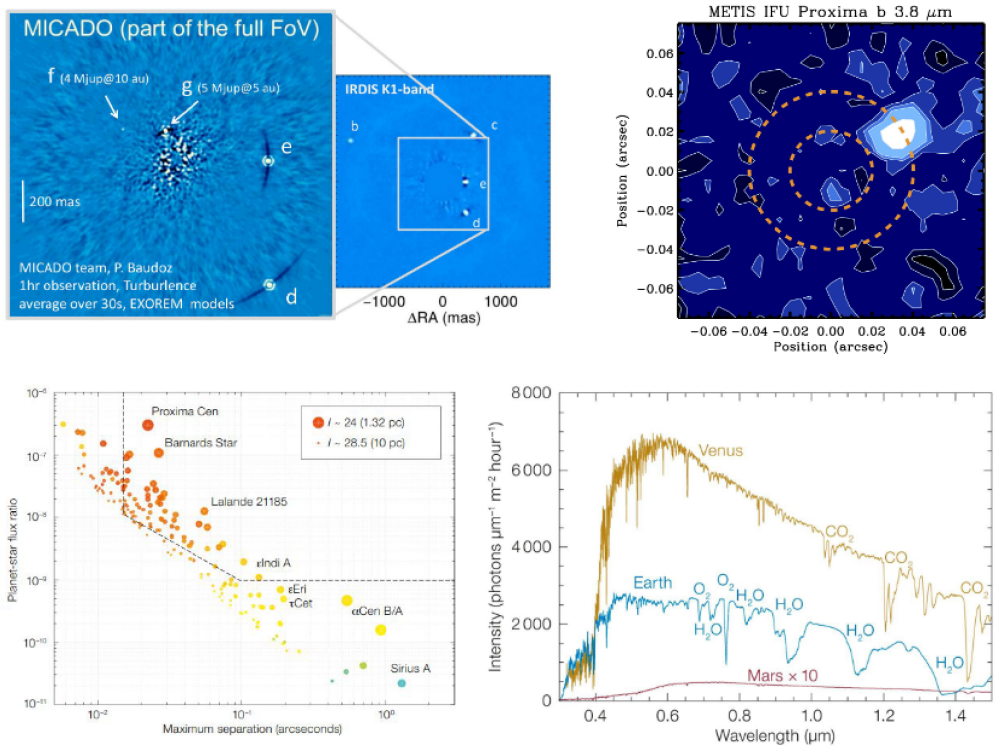

**Figure 12.** Top-Left: MICADO simulations of Direct Imaging observations of the HR 8799 planetary systems leading to the detection of putative inner planets *f* and *g* to illustrate the contrast performances already offered with the first Light instruments of ELTs (Privite communication P. Baudoz, MICDO team). Top-Right: Molecular mapping obeservations with METIS at the E-ELT of Proxima b. Bottom-Right: Angular separation and I-band flux ratio between hypothetical exoearths (Earth size and insolation, one per star) and parent stars within 10 pc (from the Hipparcos catalogue) observable from Cerro Armazones. The symbol size indicates the planet's apparent brightness, and the colours indicate stellar spectral types (red: M-stars, yellow: solar-type stars). The dotted lines indicate the approximate contrast boundaries for PCS. Bottom-Right: Spectra of Earth, Venus, and Mars the visible to near infrared wavelength range (the reflected flux of Mars has been multiplied by 10 for better visibility). Credit: ESO/Baudoz *et al.*, Absil *et al.*, Kasper *et al.*, and adapted from [90].

For very nearby planetary systems, a change of paradigm will begin by directly imaging the reflected light of exoplanets. Planet to star contrast of $10^{-5}$ to $10^{-9}$ are typically expected in reflected light for exoplanets, cold Jupiters down to super-Earths, orbiting at separation down to 10 mas. For our nearest neighbors like Alpha Cen A and Proxima, METIS and ANDES might take the first steps towards the direct detection and characterisation of temperate, rocky terrestrial planets (given their favorable planet-to-star contrast or large angular separation; see Figure 12). However, to routinely image and characterize super-Earths and Exo-Earths in the Habitable Zone (0.02 au at 10 pc around M dwarfs), the ultimate goal will be to reach the necessary contrast of about $10^{-9}$ at angular separations of 10–20 mas around nearby dwarfs with a dedicated instrument like the Planetary Camera and Spectrograph (PCS) at E-ELT, combining xAO, coronagraphy and spectroscopy. It represents a very ambitious specification directly linked with the motivation to detect bio-signatures like $O_2$, $O_3$, $CH_4$, $CO_2$ and $H_2O$ in the atmospheres of terrestrial planets in habitable zones, and possibly the first "indicative" discovery of exo-life, a revolutionary breakthrough of great scientific and philosophical value.

## Conflicts of interest

The author has no conflict of interest to declare.